\documentclass[conference,10pt,twocolumn]{IEEEtran}
\IEEEoverridecommandlockouts
\usepackage{cite}
\usepackage{amsmath,amssymb,amsfonts,amsthm}
\usepackage{graphicx}
\usepackage{textcomp}
\usepackage{xcolor}
\usepackage{dsfont}
\usepackage{bm}
\usepackage[left=0.625in, right=0.625in, top=0.7in, bottom=1.0in]{geometry}

\usepackage{subcaption}

\usepackage{algorithm}
\usepackage[noend]{algorithmic}

\newtheorem{example}{Example}
\newtheorem{theorem}{Theorem}
\newtheorem{lemma}{Lemma}

\newtheorem{definition}{Definition}

\newtheorem{remark}{Remark}

 \title{Adversary-Aware Private Inference over Wireless Channels} 
\author{
    \IEEEauthorblockN{Mohamed Seif}
    \IEEEauthorblockA{Princeton University}
    \and
    \IEEEauthorblockN{Malcolm Egan}
    \IEEEauthorblockA{Inria, France}
    \and
    \IEEEauthorblockN{Andrea J. Goldsmith}
    \IEEEauthorblockA{Stony Brook University}
    \and
    \IEEEauthorblockN{H. Vincent Poor}
    \IEEEauthorblockA{Princeton University}
}

\date{}

\begin{document}
\maketitle

\begin{abstract}
AI-based sensing at wireless edge devices has the potential to significantly enhance Artificial Intelligence (AI) applications, particularly for vision and perception tasks such as in autonomous driving and environmental monitoring. AI systems rely both on efficient model learning and inference. In the inference phase, features extracted from sensing data are utilized for prediction tasks (e.g., classification or regression). In edge networks, sensors and model servers are often not co-located, which requires communication of features. As sensitive personal data can be reconstructed by an adversary, transformation of the features are required to reduce the risk of privacy violations. While differential privacy mechanisms provide a means of protecting finite datasets, protection of individual features has not been addressed. In this paper, we propose a novel framework for privacy-preserving AI-based sensing, where devices apply transformations of extracted features before transmission to a model server. Our framework provides rigorous theoretical guarantees on the adversarial reconstruction error, which in turn offers key insights for the design of wireless channel–aware privacy mechanisms. 
\end{abstract}


\section{Introduction}

AI is expected to be a key enabler of new applications in next-generation networks \cite{saad2019vision, letaief2019roadmap, kairouz2021advances}. Combined with advanced sensing technologies, AI facilitates low-latency inference and enhance sensing applications ranging from autonomous driving to personal identification and environmental monitoring \cite{liu2020livemap, liu2020livemap}. However, sensing devices often cannot support on-device inference with high-accuracy models. On the other hand, on-server inference requires communication of raw data from sensors to an edge server, which introduces significant overheads in low-latency applications. 

A compelling alternative is collaborative inference \cite{shlezinger2022collaborative, yilmaz2024private, seif2024collaborative}, where sensing devices locally extract and communicate features to a server. In this approach, inference involves four stages: data acquisition via sensing; local feature extraction; feature encoding and transmission to a model server; and prediction at the server. As features are communicated, there is a risk of leakage of fine-grained information which must be accounted for in the design of feature encoding and transmission. 


\begin{figure*}[t]
    \centering
    \includegraphics[width= 1.4\columnwidth]{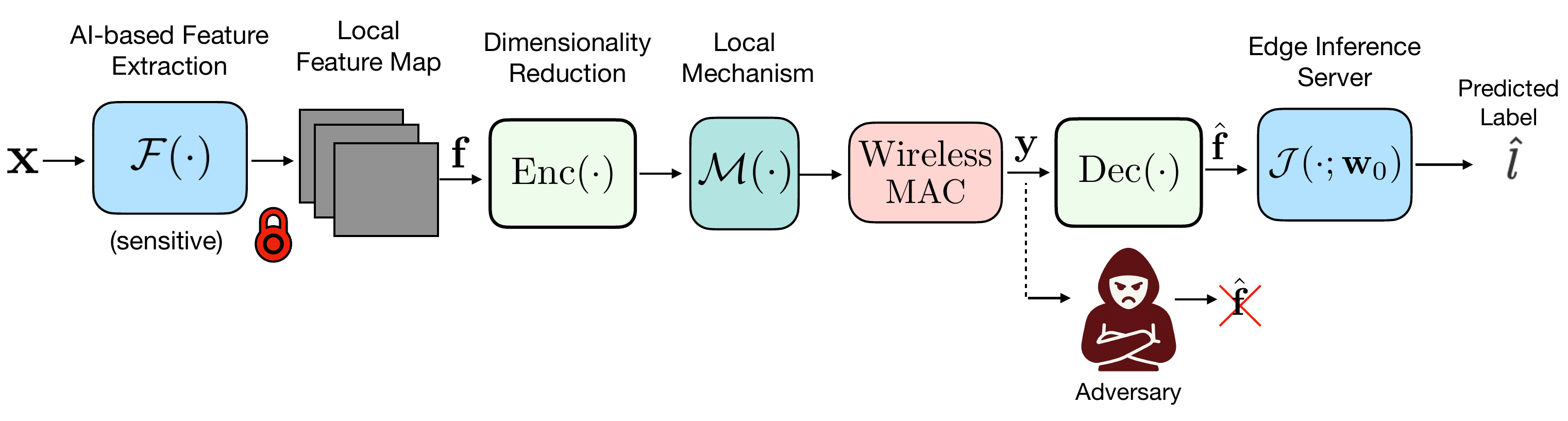}
    \caption{\small{Illustration of the private task-inference framework:  A single edge device extracts and transmits privatized features over a wireless channel to a central inference server for classification, while an adversary observes the transmission to reconstruct sensitive information.}}
    \label{fig:proposed_inference_model}
\end{figure*}

A standard approach to ensuring privacy of datasets is differential privacy, which guarantees that the probability an output of a privacy-preserving mechanism is observed does not significantly differ when an element of the dataset is removed \cite{dwork2014algorithmic}. Recently, feature differential privacy has been introduced in \cite{seif2024collaborative} for multi-user transmission of features over wireless channels. Analogous to differential privacy, the notion of feature differential privacy in \cite{seif2024collaborative} provides guarantees on the identifiability of a feature transmitted by any of the devices.

In this paper, we consider a single device which communicates encoded features to a model server over a wireless channel. At the same time, an adversary attempts to infer the communicated feature via observations over another wireless channel. The key question that we address is: \textit{how should the device encode features with a low communication overhead in order to guarantee a high reconstruction error by the eavesdropper?}   

To address this question, we introduce an end-to-end transmission pipeline for collaborative inference that jointly reduces communication overhead and ensures privacy guarantees through dimensionality reduction, controlled perturbation, and adaptive feature encoding. In contrast to privacy-agnostic communication methods that primarily emphasize reliable data delivery \cite{zhu2020toward, chen2021communication}, our approach directly integrates privacy into the transmission process by introducing calibrated randomness to mitigate reconstruction risks. Each stage of the pipeline is optimized to balance inference performance with guarantees on adversarial reconstruction error, while explicitly accounting for the impact of the wireless channel, which itself provides inherent privacy benefits. Importantly, we provide theoretical guarantees on adversarial reconstruction error, offering rigorous insights into how channel-aware feature encoding and privacy-preserving perturbation interact to safeguard sensitive information. This results in a comprehensive and efficient transmission framework for wireless AI-based sensing and inference, aligned with recent {work tailored to standard privacy criteria}  \cite{frey2021towards, yilmaz2022over, liu2023over, chen2023view, singh2023posthoc, li2024optimal, yilmaz2025private, lyu2025larger}.

\textbf{Our Contributions.} In this paper, we introduce an end-to-end pipeline for collaborative inference over wireless taking into account data acquisition and channel impairments, {beyond differential privacy constraints. In particular, we} introduce a novel privacy framework refining \cite{seif2024collaborative}, known as \textit{feature differential privacy}, designed to protect intermediate feature representations extracted at edge devices during transmission. Our main contributions are summarized as follows:
\begin{itemize}
    \item We introduce the notion of \textit{feature privacy} and establish novel theoretical guarantees against \emph{reconstruction attacks} by deriving a lower bound on the adversary’s mean squared error (MSE) as a function of model parameters, channel conditions, transmit power, and  noise.

    \item We also derive a  lower bound on the classification accuracy under differential privacy constraints, capturing the trade-offs between feature dimensions, privacy noise, and wireless channel conditions.
    
\end{itemize}

\textbf{Paper Organization.} The remainder of the paper is organized as follows. Section \ref{sec:system_model} introduces the system model and describes the proposed transmission scheme, and analyzing its classification performance in Section \ref{sec:accuracy}. Section \ref{sec:reconstruction_analysis} analyzes the adversarial MSE under the proposed scheme.  Finally, Section \ref{sec:conclusions} concludes the paper. Due to space limitations, all proofs are omitted.

\section{System Model}
\label{sec:system_model}

We consider an end-to-end transmission pipeline (Fig.~\ref{fig:proposed_inference_model}) in which a single-antenna edge device communicates its extracted features to a remote server over a wireless fading channel. Specifically, the device employs a pre-trained sub-model to process a captured signal (e.g., an image) and generate a real-valued feature vector. This representation is then transmitted to the server, where a pre-trained server-side model completes the inference by performing a classification task for {a raw input} $\mathbf{x}$ associated with a ground truth label $l^{*}$. We detail the inference procedure next.

\vspace{0.2em}
\noindent \textbf{Step 1: Feature Extraction and Dimension Reduction.}  
The device first extracts features from the raw input \(\mathbf{x}\), resulting in a vector representation \(\mathcal{F}(\mathbf{x}) \in \mathbb{R}^d\). The feature vector $\mathcal{F}(\mathbf{x})$ is then clipped via 
\begin{align}
\mathbf{f} = \min \left(1, \frac{C_{f}}{\| \mathcal{F}(\mathbf{x}) \|_{2}}\right)  \cdot \mathcal{F}(\mathbf{x}).
\end{align}
Dimension reduction is then applied using a linear encoder:
\begin{align}
\mathbf{z} = \mathbf{W} \, \mathbf{f},
\end{align}
where \(\mathbf{W} \in \mathbb{R}^{r \times d}\) is the encoder matrix (the structure of the matrix $\mathbf{W}$ will be described later), and \(r \leq d\) denotes the reduced feature dimension.
\vspace{0.2em}

\noindent \textbf{Step 2: Local Noise Injection for Privacy.}  
To ensure local privacy, the device perturbs the encoded vector {via the Gaussian mechanism \cite{dwork2014algorithmic}} by adding Gaussian noise:
\begin{align}
\tilde{\mathbf{z}} = \mathbf{z} + \mathbf{n}, \quad \mathbf{n} \sim \mathcal{N}(\mathbf{0}, \sigma^2 \mathbf{I}_r),
\end{align}
where \(\sigma^2\) is the privacy noise variance. 

\vspace{0.2em}
\noindent \textbf{Step 3: Feature Transmission.}  
The noisy feature vector $\tilde{\mathbf{z}}$ is scaled by a scaling coefficient \(\alpha > 0\) and transmitted over the wireless fading channel:
\begin{align}
\mathbf{z}' = \alpha \tilde{\mathbf{z}}, \quad \alpha = \sqrt{P},
\label{eqn:transmitted_signal}
\end{align}
where $P$ is the transmit power.

\vspace{0.2em}
\noindent \textbf{Step 4: Signal Reception at the Server.}  
The edge server receives the following signal:
\begin{align}
\mathbf{y} =  h \mathbf{z}' = h \alpha \mathbf{z} + h \alpha \mathbf{n} + \mathbf{m},
\end{align}
where \(h\) is a block-fading channel and \(\mathbf{m} \sim \mathcal{N}(\mathbf{0}, \sigma_m^2 \mathbf{I}_r)\) represents the Gaussian channel noise. We note that this model arises as a baseband representation of analog narrowband transmission (e.g., modulation of a real-valued signal via frequency-division multiplexing).

\vspace{0.2em}
\noindent \textbf{Step 5: Post-Processing.}  
The server rescales the received signal using a general scaling factor $\beta > 0$, yielding an estimate of the perturbed feature:
\begin{align}
\hat{\mathbf{z}} = \beta \mathbf{y} = \beta h \alpha \mathbf{z} + \beta h \alpha \mathbf{n} + \beta \mathbf{m}.
\end{align}

\vspace{0.2em}
\noindent \textbf{Step 6: Decoding.}  
Finally, the server reconstructs the original feature representation using a linear decoder \(\mathbf{D} \in \mathbb{R}^{d \times r}\):
\begin{align}
\hat{\mathbf{f}} = \mathbf{D} \hat{\mathbf{z}} = \beta h \alpha \mathbf{D} \mathbf{z} + \beta h \alpha \mathbf{D} \mathbf{n} + \beta \mathbf{D} \mathbf{m}.
\end{align}

\section{Server Performance}
\label{sec:accuracy}

\subsection{Classifier Accuracy}

{A key question for any private inference scheme is the classification accuracy. In order to analyze the accuracy, we impose a standard assumption on the classifier at the server. In particular,} the deployed server model has an intrinsic classification margin $\Delta$ \cite{sokolic2017robust}, formally defined as follows.



\begin{definition} [Classification Margin \cite{sokolic2017robust}] The classification margin of a target $\mathbf{x}$ represented by $( \mathbf{f}^{*}, {l}^{*})$ measured by  Euclidean distance is defined as follows:
\begin{align}
    \Delta \triangleq \sup \{B: \|{\mathbf{f}} - \mathbf{f}^{*}\|_{2} \leq B \hspace{0.1in} \operatorname{s.t.} \hspace{0.1in} \hat{l}(\mathbf{f}) =  l^{*}, \forall \mathbf{f}  \}.
\end{align}
\end{definition}

We next establish a lower bound for the classification accuracy of our proposed scheme. 

\begin{lemma} [Classification Accuracy] \label{thm:lower_bound_classification} The lower bound on the classification accuracy for our proposed privacy-preserving method can be expressed as
\begin{align}
P(\hat{l} = l^{*}) \geq \max \left\{0, P_{0} \cdot \left(1 - \frac{\operatorname{MSE}}{\Delta^{2}}\right)\right\},
\end{align}
where $P_{0}$ represents the classification accuracy in the ideal case (i.e., no communication errors and privacy constraints), and $\Delta$ represents the inherent classification margin.
\end{lemma}

\subsection{Upper bound on the MSE at the Server}

We analyze the degradation in inference accuracy induced by local privacy noise and wireless channel impairments. Specifically, we quantify the MSE between the original feature vector \(\mathbf{f} \in \mathbb{R}^d\) extracted at the edge device and the reconstructed estimate \(\hat{\mathbf{f}} \in \mathbb{R}^d\) at the server. The MSE at the server is given by:
\[
\operatorname{MSE} \triangleq \mathbb{E} \left[ \| \hat{\mathbf{f}} - \mathbf{f} \|_2^2 \right],
\]
which can be upper bounded as:
\begin{align}
\operatorname{MSE} 
&\leq \left\| \beta h \alpha \mathbf{D} \mathbf{W} - \mathbf{I}_d \right\|_F^2 \cdot \| \mathbf{f} \|_2^2 \nonumber \\
&\quad + \beta^2 h^2 \alpha^2 \sigma^2 \cdot \| \mathbf{D} \|_F^2 
+ \beta^2 \sigma_m^2 \cdot \| \mathbf{D} \|_F^2.
\end{align}

\noindent This expression consists of three key components:
\begin{enumerate}
    \item[(i)] {approximation error} due to the deviation of \(\beta h \alpha \mathbf{D} \mathbf{W}\) from the identity matrix, which reflects the encoder-decoder mismatch. {As $\mathbf{W}$ has linearly independent columns with probability one, choosing $\mathbf{D}$ as the Moore-Penrose pseudoinverse allows reconstruction of $\mathbf{z}$ in the absence of noise. As a consequence if $\beta = \frac{1}{\alpha h}$, the approximation error can be zero. However, this requires perfect estimation of $h$.} 
    \item[(ii)] {privacy-induced distortion} that is arising from the locally injected Gaussian noise \(\mathbf{n}\), scaled by transmission and channel gain. 
    \item[(iii)] {channel-induced distortion} resulting from the additive Gaussian channel noise \(\mathbf{m}\). In addition, the expectation is taken over both sources of noise, while the feature vector \(\mathbf{f}\) is assumed deterministic after clipping, satisfying \(\|\mathbf{f}\|_2 \leq C_f\).
\end{enumerate}

\begin{figure}[t]
	\centering
    {\includegraphics[width=0.4\textwidth]{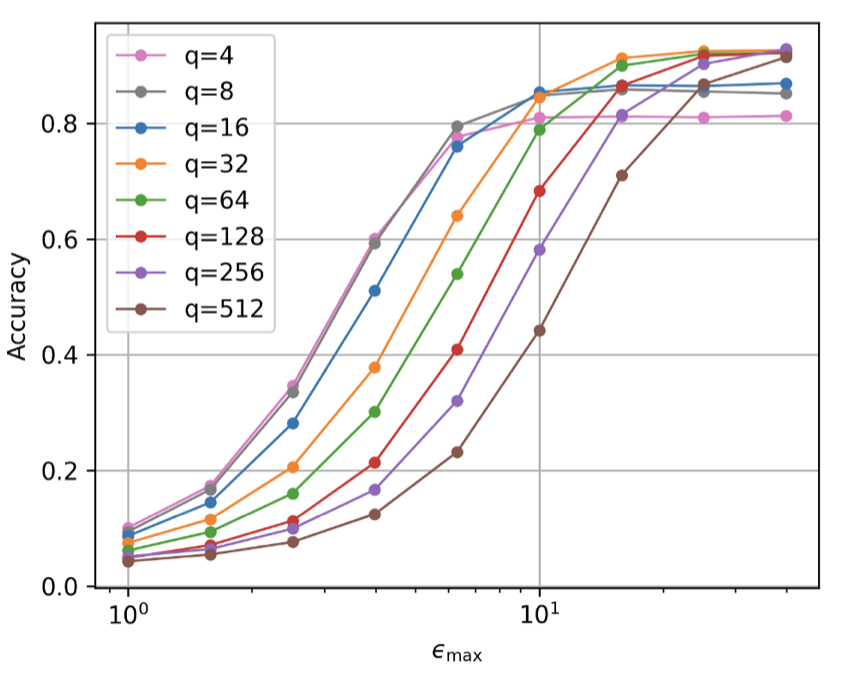}}
    \caption{\small{Impact of the dimensionality reduction on the classification accuracy for the same privacy leakage $\epsilon_{\max}$, where $r = q \times 7 \times 7$, based on the ModelNet Dataset \cite{wu20153d}, where $P = 30$ dBm, $\sigma^{2} = 0.1$, $\delta = 10^{-5}$, and $C_{f} = 10^{2}$.}}
    \label{fig:impact_of_hete_dimensionality}
    \vspace{-10pt}
\end{figure}

In high-dimensional feature spaces, the performance of the model can degrade due to the increasing impact of perturbation noise. Specifically, {{in the \textit{high} privacy regime,}} the noise introduced for privacy protection scales with the dimensionality of the data, causing the MSE to increase. The MSE behaves asymptotically as: $\operatorname{MSE} \sim O(d)$, where \(d\) represents the dimensionality of the feature space \cite{seif2024collaborative}. The classification accuracy bound then gives:
\[
P(\hat{l} = l^{*}) \geq P_0 \left(1 - \frac{O(d)}{\Delta^2}\right).
\]
As \(d \to \infty\), the classification accuracy degrades, approaching zero unless the classification margin \(\Delta\) grows to counterbalance the increasing MSE. Therefore, while higher dimensions provide more expressive power for feature representation, they also amplify the effects of perturbation noise, leading to a reduction in accuracy. We show this impact in Fig. {{\ref{fig:impact_of_hete_dimensionality}}, which demonstrates how classification accuracy decreases as dimensionality increases, driven by the amplification of perturbation noise. {{However, in the \textit{low} privacy regime, the relationship between dimensionality and accuracy is different \cite{seif2024collaborative}. In this regime, the impact of noise is not dominant. This effect is illustrated in Fig. 2, where the accuracy for the case when \(q = 16\) starts to degrade at higher values of \(q\) for \(\epsilon \geq 10\), indicating a critical turning point where additional dimensionality begins to aid rather than impede classification accuracy.
}}

\section{Reconstruction Attack Analysis}
\label{sec:reconstruction_analysis}

In this section, we analyze the privacy leakage from an adversary attempting to reconstruct the encoded feature vector \(\mathbf{z}\) by eavesdropping on the wireless transmission over another channel.

\vspace{0.5em}
\noindent \textbf{Adversarial Observation.}  The adversary observes
\begin{align}
\mathbf{y}_{\mathrm{adv}} = g \alpha \mathbf{z} + g \alpha \mathbf{n} + \mathbf{m}_{\mathrm{adv}}, \label{eqn:adversarial_observation}
\end{align}
where the {adverary's channel is block fading with channel coefficient \(g \in \mathbb{R}\)}, and \(\mathbf{m}_{\mathrm{adv}} \sim \mathcal{N}(\mathbf{0}, \sigma_a^2 \mathbf{I}_r)\) denotes the additive Gaussian noise at the adversary’s receiver. {We define the total effective noise variance of the adversary as}
\[
\nu^2 = g^2 \alpha^2 \sigma^2 + \sigma_a^2.
\]

\vspace{0.2em}
\noindent \textbf{Generic Linear Estimator.}  
The adversary applies a re-scaling factor \(\gamma > 0\) to form an estimate of the latent representation:
\begin{align}
\hat{\mathbf{z}} = \gamma \mathbf{y}_{\mathrm{adv}} = \gamma g \alpha \mathbf{z} + \gamma g \alpha \mathbf{n} + \gamma \mathbf{m}_{\mathrm{adv}}.
\end{align}
Unless \(\gamma = \frac{1}{g \alpha}\), the estimator is biased. {As the adversary will also have channel estimation errors, we retain \(\gamma\) as a tunable parameter dependent on the estimation error to explore trade-offs between bias and variance.}

\vspace{0.2em}
\noindent \textbf{Reconstruction Error.}  
The MSE reconstruction error at the adversary is given as
\begin{align}
& \mathrm{MSE}_{\mathrm{adv}}  = \mathbb{E} \left[ \| \hat{\mathbf{z}} - \mathbf{z} \|^2 \right] \nonumber \\
& = (\gamma g \alpha - 1)^{2} \left\| \mathbf{z} \right\|^2 + \gamma^2 g^2 \alpha^2 \mathbb{E}[\| \mathbf{n} \|^2] + \gamma^2 \mathbb{E}[\| \mathbf{m}_{\mathrm{adv}} \|^2]. \nonumber
\end{align}
We next provide a formal privacy definition for protecting the transmitted signals, followed by a detailed analysis of the resulting adversarial error.

\begin{definition} 
[$(\epsilon, \delta)$-feature DP] {Let $\mathbf{x}$ be a raw input and $\mathcal{F}$ be an associated space of features and $C_f > 0$.}  A randomized mechanism $\mathcal{M}: \mathcal{F} \rightarrow \mathds{R}^{d}$ is $(\epsilon, \delta)$-feature DP if for any two features $\mathbf{f}, \mathbf{f}' \in \mathcal{F}$ satisfying $\|\mathbf{f} - \mathbf{f}'\| \leq 2C_f$, and any measurable subset $\mathcal{S} \subseteq \text{Range}(\mathcal{M})$, we have
\begin{align}
    \operatorname{Pr}(\mathcal{M}(\mathbf{f}) \in \mathcal{S}) \leq e^{\epsilon} \operatorname{Pr}(\mathcal{M}(\mathbf{f}') \in \mathcal{S}) + \delta.
\end{align}
The setting when $\delta = 0$ is referred as pure $\epsilon$-feature DP. 
\end{definition}


We next present a lower bound on the adversarial MSE for the case when the elements of the compression matrix $\mathbf{W}$ are drawn from Laplacian distribution. 
\begin{theorem}[Adversarial MSE Lower Bound under DP]
\label{thm:adv-mse}
Let \( \mathbf{f} \in \mathbb{R}^d \) be a clipped feature satisfying \( \| \mathbf{f} \|_2 \le C_f \). Let \( \mathbf{z} = \mathbf{W} \mathbf{f} \in \mathbb{R}^r \), where \( \mathbf{W} \in \mathbb{R}^{r \times d} \) is a random matrix with i.i.d. entries drawn from the Laplace distribution \( \mathrm{Lap}(0, b) \). 
Then, with probability at least \( 1 - \delta \) over the draw of \( \mathbf{W} \), the mechanism 
\[
\mathcal{M}(\mathbf{f}) = \mathbf{W} \mathbf{f} + \mathbf{n}
\]
satisfies \((\epsilon, 2 \delta)\)-feature DP with respect to the input feature vector \( \mathbf{f} \), provided
\[
\sigma^2 = \frac{8 C_w^2 b^2 C_f^2 (r + d) \log(1.25/\delta)}{\epsilon^2},
\]
where \( C_w = 4 \left( 1 + \frac{\log(2/\delta)}{\sqrt{r} + \sqrt{d}} \right) \) is a high-probability upper bound on the spectral norm \( \| \mathbf{W} \|_2 \). 
Moreover, 
\[
\mathrm{MSE}_{\mathrm{adv}} \geq \frac{r \nu^2 D^2}{g^2 \alpha^2 D^2 + r \nu^2}, \quad D = C_{f} C_{w} b (\sqrt{r} + \sqrt{d}).
\]
\end{theorem}
We visualize the impact of the transmitted signal dimension 
$r$ on the adversarial MSE in Fig.~\ref{fig:mse-eps}. While a smaller  $r$ typically reduces communication latency, it also limits the adversary’s reconstruction error, potentially reducing privacy leakage.


\vspace{0.5em}
\noindent \textit{Proof Sketch.} To ensure \((\epsilon, 2 \delta)\)-feature DP as presented in the above Theorem, we use the Gaussian mechanism \cite{dwork2014algorithmic}, which requires the noise variance to scale with the squared global \(\ell_2\)-sensitivity of the mechanism. For neighboring features with \(\| \mathbf{f} - \mathbf{f}' \|_2 \le 2 C_f\), the sensitivity of the linear mapping \(\mathbf{W} \mathbf{f}\) is
\[
\Delta_2 = \| \mathbf{W} (\mathbf{f} - \mathbf{f}') \|_2 \le \| \mathbf{W} \|_2 \cdot \| \mathbf{f} - \mathbf{f}' \|_2 \le 2 C_f \cdot \| \mathbf{W} \|_2.
\]
With high probability \(1 - \delta\), the spectral norm {of $\mathbf{W}$} is bounded by \( C_w b (\sqrt{r} + \sqrt{d}) \) \cite{blocki2012johnson}. Hence,
\[
\Delta_2 \le 2 C_f C_w b (\sqrt{r} + \sqrt{d}).
\]
By the Gaussian mechanism \cite{dwork2014algorithmic}, it suffices to use
\[
\sigma^2 \ge \frac{2 \Delta_2^2 \log(1.25/\delta)}{\epsilon^2}
= \frac{8 C_w^2 b^2 C_f^2 (\sqrt{r} + \sqrt{d})^2 \log(1.25/\delta)}{\epsilon^2}.
\]
Using \( (\sqrt{r} + \sqrt{d})^2 \le 2(r + d) \), we further simplify:
\[
\sigma^2 = \frac{8 C_w^2 b^2 C_f^2 (r + d) \log(1.25/\delta)}{\epsilon^2}.
\]
Now consider the adversarial observation:
\[
\mathbf{y}_{\mathrm{adv}} = g \alpha \mathbf{z} + g \alpha \mathbf{n} + \mathbf{m}_{\mathrm{adv}} = g \alpha \mathbf{z} + \mathbf{w},
\]
where \( \mathbf{w} \sim \mathcal{N}(0, \nu^2 \mathbf{I}_r) \) and \( \nu^2 = g^2 \alpha^2 \sigma^2 + \sigma_a^2 \).

We consider a general linear estimator \( \hat{\mathbf{z}} = \gamma \mathbf{y}_{\mathrm{adv}} \). The minimax MSE over all \( \mathbf{z} \) satisfying \( \|\mathbf{z}\|_2^2 \le D^2 \) is
\[
\mathrm{MSE}_{\mathrm{adv}} = \inf_{\gamma} \sup_{\|\mathbf{z}\|_2 \le D}  \left[ (\gamma g \alpha - 1)^2 \|\mathbf{z}\|_2^{2} + \gamma^2 r \nu^2 \right].
\]
Solving for the optimal \( \gamma^* = \frac{g \alpha D^2}{g^2 \alpha^2 D^2 + r \nu^2} \) and substituting back, we obtain
\[
\mathrm{MSE}_{\mathrm{adv}} \geq \frac{r \nu^2 D^2}{g^2 \alpha^2 D^2 + r \nu^2}.
\]

We next determine the optimal projection dimension $r$ required to ensure a target adversarial MSE threshold 
$\Omega$, while minimizing latency. The result is given below.
\begin{lemma}[Optimal Projection Dimension under Adversarial MSE Constraint]
\label{thm:optimal-r-alpha-power}
Then, the minimal encoding dimension \( r^* \in \mathbb{N} \) that minimizes latency and satisfying the constraint $\mathrm{MSE}_{\mathrm{adv}} \geq \Omega$, it suffices to choose $r$ as 
\begin{align}
r^{*} = \left\lceil \frac{g^2 \alpha^2 D^2 \cdot \Omega}{\nu^2 \left(D^2 - \Omega\right)} \right \rceil.
\end{align}
\end{lemma}
\noindent After simplifications (up to constants), we have
\[
r^{\star}
=\Theta\!\left(
\frac{
-\bigl(\tfrac{d}{\epsilon^{2}}+\sigma_{a}^{2}\bigr)
+\sqrt{\bigl(\tfrac{d}{\epsilon^{2}}+\sigma_{a}^{2}\bigr)^{2}
+\tfrac{1}{\epsilon^{2}}}
}{
\tfrac{1}{\epsilon^{2}}
}
\right),
\]
which can be further simplified order-wise in two main regimes:
\[
r^{\star} =
\begin{cases}
\Theta\!\bigl(\epsilon^{2}/d\bigr), &
\text{if } \tfrac{d}{\epsilon^{2}} \gg \sigma_{a}^{2}
\quad (\text{privacy-limited}),\\[4pt]
\Theta(1), &
\text{if } \sigma_{a}^{2} \gg \tfrac{d}{\epsilon^{2}}
\quad (\text{noise-limited}).\\
\end{cases}
\]
To summarize, stronger privacy (smaller~$\epsilon$) or larger~$d$
reduce the required projection dimension roughly as
$r^{\star}\!\propto\!\epsilon^{2}/d$,
while in the noise-limited regime $r^{\star}$ saturates to a constant
set by the channel noise~$\sigma_{a}^{2}$.

\begin{figure}[t]
    \centering
    \includegraphics[width=0.4\textwidth]{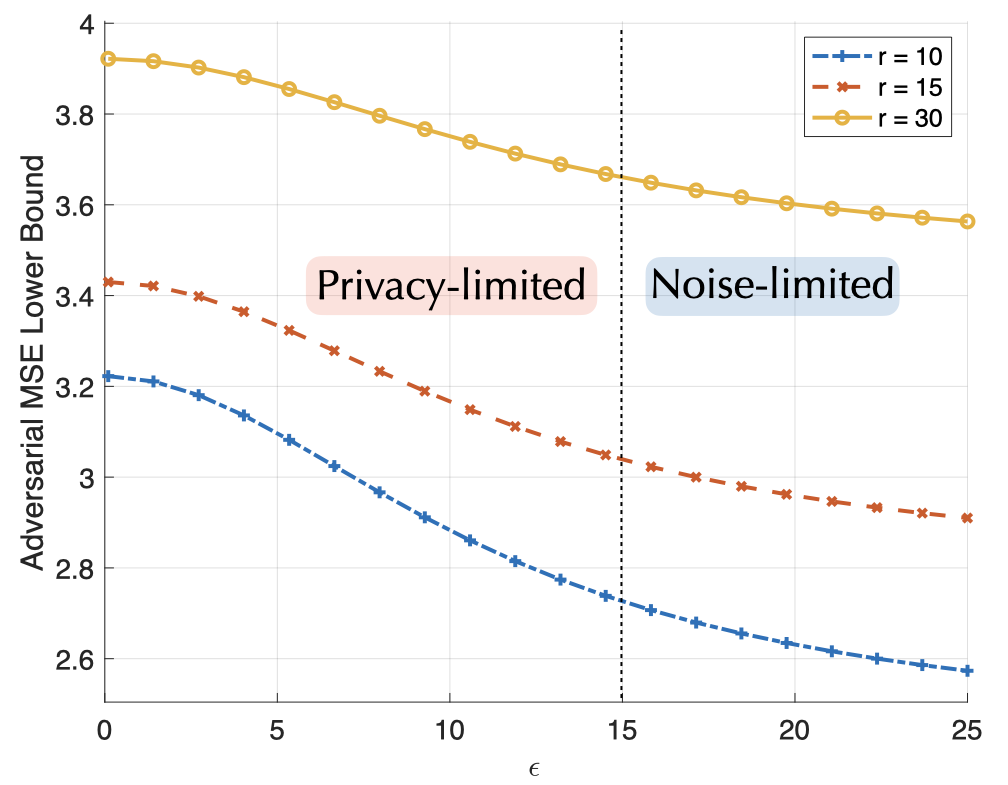} 
    \caption{\small{Adversarial MSE lower bound as a function of the privacy parameter \( \epsilon \). Parameters used: \( C_f = 2 \), \( b = 0.01 \), \( d = 50 \), \( \delta = 10^{-5} \), \( \alpha = 1 \), \( g = 1 \), and \( \sigma_a^2 = 1 \). The MSE bound increases with stronger privacy (smaller \( \epsilon \)), reflecting the privacy-utility trade-off.
    }}
    \label{fig:mse-eps}
\end{figure}

\begin{theorem}[Lower bound on adversarial MSE for recovering $\mathbf f(\mathbf x)$]
\label{thm:lower_bound_feature_mse}
Let $\hat{\mathbf z}$ be the adversary’s MMSE estimator of $\mathbf z$, and suppose the adversary forms
\[
\hat{\mathbf f}(\mathbf x)=\mathbf W^\dagger \hat{\mathbf z}, 
\qquad 
\mathbf W^\dagger := (\mathbf W^\top \mathbf W)^{-1}\mathbf W^\top,
\]
where $\mathbf W\in\mathbb R^{r\times d}$ has full row rank (so $d\ge r$). Let $\mathbf e:=\hat{\mathbf z}-\mathbf z$.
Then
\[
\hat{\mathbf f}(\mathbf x)-\mathbf f(\mathbf x)=\mathbf W^\dagger \mathbf e.
\]
Conditioning on $\mathbf W$, we have
\begin{align}
\mathbb E\!\left[\|\hat{\mathbf f}(\mathbf x)-\mathbf f(\mathbf x)\|_2^2 \mid \mathbf W\right]
&= \mathbb E\!\left[\|\mathbf W^\dagger \mathbf e\|_2^2\right] \nonumber\\
&\ge \sigma_{\min}^2(\mathbf W^\dagger)\, \cdot \mathbb E\!\left[\|\mathbf e\|_2^2\right] \nonumber\\
&= \frac{1}{\sigma_{\max}^2(\mathbf W)}\, \cdot \mathrm{MSE}_{\mathrm{adv}}.
\label{eq:lb_feature_mse}
\end{align}
Now assume $\mathbf W$ has i.i.d.\ zero-mean, unit-variance Laplace entries (more generally, i.i.d.\ isotropic
sub-exponential entries). Then there exist absolute constants $c_{0},c>0$ such that for any $t>0$,
with probability at least $1-2e^{-c t^2}$,
\begin{equation}
\sigma_{\max}(\mathbf W)\le c_0 \big(\sqrt d+\sqrt r+t\big).
\label{eq:smax_bound}
\end{equation}
Combining \eqref{eq:lb_feature_mse} and \eqref{eq:smax_bound} yields: with probability at least $1-2e^{-c t^2}$,
\[
\mathbb E\!\left[\|\hat{\mathbf f}(\mathbf x)-\mathbf f(\mathbf x)\|_2^2 \right]
\ge
\frac{1}{c_0^2\big(\sqrt d+\sqrt r+t\big)^2}\, \cdot \mathrm{MSE}_{\mathrm{adv}}.
\]
\end{theorem}

\vspace{-0.1in}
\begin{remark}[Pseudoinverse vs. Neural Network Estimators]
While a neural network could be used by the adversary to estimate \( \mathbf{f}(\mathbf{x}) \) from \( \hat{\mathbf{z}} \), the pseudoinverse remains optimal in the worst-case setting without a prior on \( \mathbf{f}(\mathbf{x}) \). It minimizes the reconstruction error among all linear estimators, and no nonlinear method can fundamentally outperform it unless additional structure or data is available. Thus, the lower bound holds for any adversarial estimator, including deep networks.
\end{remark}


\section{A Wireless Data Acquisition Model}

We next present a concrete example that incorporates a realistic data acquisition model based on noisy subsampled transform measurements. Such acquisition schemes arise in practical systems including compressive imaging, wireless spectrum sensing, and embedded hardware, where only a subset of transform-domain coefficients can be measured due to sensing, power, or bandwidth constraints. To capture this structure, we model the feature extractor as a subsampled orthogonal transform with additive measurement noise and apply clipping to ensure bounded sensitivity. This setup allows us to instantiate the general reconstruction bounds in a setting that reflects real-world constraints while preserving analytical tractability.

\vspace{0.4em}
\begin{example}[Subsampled Orthogonal Feature Acquisition]
Let the feature extractor be defined as
\[
\mathbf{f}(\mathbf{x}) = \mathbf{A} \mathbf{x} + \mathbf{w}, \quad \text{where } \mathbf{A} = \mathbf{P}_d \mathbf{T}_m \in \mathbb{R}^{d \times m},
\]
where \( \mathbf{T}_m \in \mathbb{R}^{m \times m} \) a unitary transform (e.g., DFT), so that \( \mathbf{T}_m^\top \mathbf{T}_m = \mathbf{I}_m \),  \( \mathbf{P}_d \in \{0,1\}^{d \times m} \) a subsampling operator selecting \( d \) rows uniformly without replacement, and \( \mathbf{w} \sim \mathcal{N}(0, \sigma_w^2 \mathbf{I}_d) \) is an additive measurement noise. The inverse map used by the adversary is
\[
\hat{\mathbf{x}} = \mathbf{T}_m^\top \mathbf{P}_d^\top \hat{\mathbf{f}}(\mathbf{x}),
\]
where \( \mathbf{P}_d^\top \) zero-fills the unobserved transform coordinates. Since both \( \mathbf{P}_d^\top \) and \( \mathbf{T}_m^\top \) are norm non-expanding, the inverse map is 1-Lipschitz:
\[
\| \hat{\mathbf{x}} - \mathbf{x} \|_2 \ge \| \hat{\mathbf{f}}(\mathbf{x}) - \mathbf{f}(\mathbf{x}) \|_2.
\]

Substituting the adversarial feature reconstruction bound from the previous theorem, we obtain
\[
\mathbb{E} \left[ \| \hat{\mathbf{x}} - \mathbf{x} \|_2^2 \right] \ge \frac{1}{c_{0}^2 ( \sqrt{d} + \sqrt{r} + t )^2} \cdot \mathrm{MSE}_{\mathrm{adv}},
\]
with high probability over the draw of the encoder \( \mathbf{W}  \), and where \( \mathrm{MSE}_{\mathrm{adv}} = \mathbb{E}[\| \hat{\mathbf{z}} - \mathbf{z} \|_2^2] \) is the adversary’s MMSE on the encoded representation \( \mathbf{z} \).
\end{example}


\section{Privacy Amplification via Massive MIMO.}

To further strengthen privacy at the physical layer, we extend the communication model to a massive MIMO setting where the transmitter (i.e., the client device) is equipped with a large antenna array and a single antenna receiver (i.e., the server). In this regime, the uplink channel to the server exhibits \emph{channel hardening} and \emph{favorable propagation} \cite{ngo2014aspects}, concentrating tightly around its mean, whereas the adversary’s channel remains weak and statistically uncorrelated owing to favorable propagation conditions. This asymmetry enables the legitimate receiver to coherently decode the transmitted features while inherently suppressing information leakage to potential eavesdroppers.

Under the same transmission procedure as described in Section \ref{sec:system_model}, the input–output relationship at channel use $i$ are given by
\begin{align}
\mathbf{y}[i] &= \mathbf{h}\, \boldsymbol{z}'[i] + \mathbf{m}[i], \nonumber\\
\mathbf{y}_{\mathrm{adv}}[i] &= \mathbf{h}_{\mathrm{adv}}\, \boldsymbol{z}'[i] + \mathbf{m}_{\mathrm{adv}}[i],\quad i = 1,2,\dots,r,
\end{align}
where $\mathbf{h} \in \mathbb{R}_{+}^{M}$ denotes the legitimate block-fading channel to the inference server, $\mathbf{h}_{\mathrm{adv}} \in \mathbb{R}_{+}^{M}$ represents the adversarial channel, and $\mathbf{m}[i]$, $\mathbf{m}_{\mathrm{adv}}[i]$ denote additive noise terms.  The adversary reconstructs the transmitted feature vector at channel use~$i$ using a linear estimator, i.e.,
\begin{align}
\hat{\mathbf{z}}[i] = \frac{\mathbf{h}_{\mathrm{adv}}^{\!\top}}{\alpha}\, \cdot  \mathbf{y}_{\mathrm{adv}}[i],
\end{align}
where $\alpha > 0$ is a scaling coefficient. Following the same analytical procedure as in the single-antenna case, the next lemma characterizes the corresponding adversarial reconstruction error under massive MIMO with channel hardening.

\begin{lemma}[Adversarial Reconstruction Error under Massive MIMO]
\label{lem:adv-mmse-massive-mimo}

Under the massive MIMO setting and our proposed transmission scheme, the adversary’s MMSE in estimating $\mathbf{z}$ satisfies
\begin{align}
\mathrm{MSE}_{\mathrm{adv}}
\;\ge\;
\frac{r\!\left(\tfrac{\alpha^2\sigma^2}{M} + \sigma_a^2\right)}
     {\tfrac{\alpha^2(C_z^2+\sigma^2)}{M} + \sigma_a^2}.
\end{align}
\end{lemma}
It is worth highlighting that in the massive MIMO limit $(M \to \infty)$, the lower bound converges to
$\mathrm{MSE}_{\mathrm{adv}} \to r$, indicating that the adversary asymptotically gains no information about $\mathbf{z}$.

\section{Conclusions \& Future Work} \label{sec:conclusions}

In this paper, we have introduced a new framework for adversary-aware private inference over wireless channels, termed \textit{feature} DP, which aims to protect extracted features during transmission from reconstruction attacks. We derived fundamental lower bounds on adversarial reconstruction error, highlighting how key system parameters—such as encoder structure, privacy noise, and channel noise—impact the difficulty of input recovery. As a direction for future work, we plan to explore more structured data acquisition models, to further bridge theory and practical sensing constraints. Also considering non-linear dimensionality reduction mechanisms such convolutional neural networks and investigating the fundamental tradoeff between model complexity and privacy leakage is of another direction of great interest.

\bibliographystyle{IEEEtran}
\bibliography{myreferences}

\end{document}